# The sinking dynamics and splitting of a granular droplet

Jens P. Metzger, [1,*] Christopher P. McLaren,[1,*] Sebastian Pinzello,[1] Nicholas A. Conzelmann,[1] Christopher M. Boyce,[2] Christoph R. Müller,[1,†]

[1]*Department of Mechanical and Process Engineering, ETH Zurich, 8092 Zurich, Switzerland*

[2]*Department of Chemical Engineering, Columbia University, New York, New York 10027, USA*

Recent experimental results have shown that vibro-fluidized, binary granular materials exhibit Rayleigh-Taylor-like instabilities that manifest themselves in rising plumes, rising bubbles and the sinking and splitting of granular droplets. This work explores the physics behind the splitting of a granular droplet that is composed of smaller and denser particles in a bed of larger and lighter particles. During its sinking motion, a granular droplet undergoes a series of binary splits resembling the fragmentation of a liquid droplet falling in a miscible fluid. However, different physical mechanisms cause a granular droplet to split. By applying particle-image-velocimetry and numerical simulations, we demonstrate that the droplet of high-density particles causes the formation of an immobilized zone underneath the droplet. This zone obstructs the downwards motion of the droplet and causes the droplet to spread and ultimately to split. The resulting fragments sink at inclined trajectories around the immobilized zone until another splitting event is initiated. The occurrence of consecutive splitting events is explained by the re-formation of an immobilized zone underneath the droplet fragments. Our investigations identified three requirements for a granular droplet to split: 1) frictional inter-particle contacts, 2) a higher density of the particles composing the granular droplet compared to the bulk particles and 3) and a minimal granular droplet diameter.

## I. INTRODUCTION

Granular motion leads to a series of natural phenomena such as sand dunes [1], rock slides [2] or avalanches [3]. In addition, a variety of industries, including the pharmaceutical and food industries [4] rely on granular materials and the handling thereof to manufacture their desired products. Granular materials are complex as they can exhibit solid-like, liquid-like or gas-like properties depending on the degree of external agitation [5]. One example is a fluid flowing upwards through a particle packing. At low fluid velocities, the packing is stationary and solid-like, yet if the fluid velocity is faster than the so-called minimum fluidization velocity ($U_{mf}$), the drag forces overcome gravity and the granular particles are in a fluidized state, i.e. the granular material exhibits liquid-like properties such as the rise of gas bubbles [6-9] in the bulk phase.

A prominent example of a classic fluid phenomena is the Rayleigh-Taylor (RT) instability in which dense and light fluids mix due to differences in buoyancy [10]. When two fluids of different density are layered, with the denser fluid on top of the lighter one, the RT instability creates interpenetrating "fingers" at the fluid interphase. Such fingering patterns, reminiscent of RT instabilities have been observed very recently also in granular media [11,12]. For example, RT-like patterns have been found to occur when a closed container is filled with either a liquid [13] or a gas [14] and the container is suddenly inverted, bringing previously sedimented particles to the top of the container. More recently, McLaren et al. [15] observed RT-like instabilities in a binary set of dry granular materials that are simultaneously mobilized by fluidizing air and vibration. In the work of McLaren et al. [15], a layer of larger and lighter particles below a layer of smaller and denser particles forms interpenetrating fingers.

---

[*] Shared first authors
[†] Corresponding author: muelchri@ethz.ch

The RT-like instability discovered by McLaren et al. [15] gives rise to another phenomenon that is typically also observed in fluids, i.e. the fragmentation of a liquid droplet falling in another miscible liquid [16]. In miscible fluids, a dense liquid droplet falling in a miscible but less dense liquid forms a torus. This torus increases in diameter until the RT instability creates discrete circumferential aggregations that become large enough to separate and form new toroidal structures [16-20]. Such fragmentation instabilities have also been observed in two-phase granular materials. When droplets of a suspension of glycerol-water and glass spheres sink in a glycerol-water solution at very low Reynolds numbers, the droplet transforms into a ring with increasing radius [21]. In some cases, these droplets even exhibit a ring instability as seen in pure liquids [22]. McLaren et al. [15] demonstrated the presence of a fragmentation instability in dry granular materials when a so called "granular droplet" falls in another granular material. A granular droplet is an aggregation of smaller particles with diameter $d_d$ and higher density $\rho_d$ surrounded by a bulk of particles with a larger diameter $d_b$ but lower density $\rho_b$. When the granular system is agitated by a fluidizing gas flow and vertical vibration, the granular droplet sinks through the bulk phase. Instead of sinking downwards in a straight motion, the granular droplet splits into two daughter droplets, which descend along inclined paths whereby the daughter droplets formed repeat the process. The structures observed show a striking similarity to a vertical cross section through a ring fragmentation formed by liquid droplets. However unlike liquids, granular particles interact via frictional contact forces which can readily solidify the granular material and counteract a fluid-like motion, unless the particles are sufficiently agitated [5]. Therefore, the underlying mechanism driving granular droplet splitting must be different from the RT instability found in their liquid counterpart.

McLaren et al. [15] argued that a heterogeneous gas flow pattern forms around a granular droplet due to size differences between the granular droplet particles and bulk particles. They hypothesized this heterogeneity to cause locally solidified regions that initiate a droplet to split. However, gas heterogeneities alone cannot fully explain the dynamics of the splitting phenomenon, e.g. why the fragments formed during the first split decent further without splitting and then suddenly perform another split. Furthermore, it is unclear, if any fragmentation occurs when the gas flows homogeneously through the droplet and the surrounding bulk phase. Therefore, this study aims to elucidate the detailed mechanisms driving the motion and splitting of a granular droplet. In addition, we will determine the droplet properties that are necessary for a granular droplet to split. To this end, we combined experimental investigations applying a high-speed camera and particle image velocimetry (PIV) and numerical simulations to probe the effect of contact forces and gas flow dynamics.

## II. METHODS

### A. Experimental setup, digital image analysis and particle image velocimetry

A pseudo-two-dimensional fluidized bed with a width of 400 mm, a height of 500 mm and a transverse thickness of 10 mm was used. The walls were made of polymethyl methacrylate (PMMA) sheets. Air for fluidization entered the bed through 40 evenly spaced holes of diameter 1.1 mm drilled into a PMMA distributor plate. The granular material consisted of two sets of spherical particles (Sigmund Lindner GmbH, SiLibeads): (i) Black colored glass particles with density $\rho_b$ = 2500 kg/m³ and mean diameter $d_b$ = 1.7 mm (referred to as bulk particles, index b) and (ii) white zirconium oxide particles with density $\rho_d$ = 6000 kg/m³ and mean diameter $d_d$ = 1.1 mm (referred to as droplet particles, index d), see TABLE I. The bulk and droplet particles were chosen to have close minimum fluidization velocities in the non-vibrated state ($U_{mf}$). $U_{mf}$ of the bulk particles is 0.9 m/s. The initial particle packing was prepared by first filling the bed with the denser bulk particles to a height of 410 mm. The granular droplet was constructed on top of the bulk particles with its bottom edge located at a height of 410 mm. For this, two thin retaining walls were inserted into the freeboard of the bed dividing it into three vertical compartments. The central compartment had a width of $D$ and was filled with droplet particles to a height $D$ to construct a square-cuboid granular droplet of edge length $D$. Bulk particles were added on top of the granular droplet and in the outer two compartments to yield a final packing height of 460 mm. Once all three compartments were filled, the retaining walls were removed. By varying the distance between the two retaining walls as well as the height $D$ to which the droplet particles were poured in, granular droplets of variable edge length ($D$ = 10, 20, 30, 40, 50 mm) were constructed. The first panel in FIG. 1 shows the initial state of a bed containing a granular droplet with $D$ = 50 mm. The experiment was started by simultaneously vibrating and fluidizing the bed. To impose a defined vibration on the system a Labworks shaker (Labworks Inc., ET-139) was used generating a vertical sinusoidal displacement $\delta = A \sin(2\pi f t)$. Besides the shaker, the vibration system contains a controller (Labworks Inc. VL-144), an amplifier (Labworks Inc., PA-138-1), and an accelerometer (PCB Piezotronics Inc., J352C33.). In the experiments, the amplitude $A$ and frequency $f$ were varied in the range of $A$ = [0.05 mm, 1 mm] and $f$ = [10 Hz, 47.4 Hz]. The frequency and amplitudes used, resulted in a span of dimensionless vibration strengths $\Gamma = A(2\pi f)^2/g$ = [0.18, 0.45], where $\Gamma$ is the peak acceleration of the vibration normalized by the magnitude of the gravitational acceleration $g$. The fluidization airflow was controlled using a mass flow controller (Bronkhorst AG, F-203AV). After the mass flow controller, the air passed through a humidifier to ensure a constant relative humidity of 87-91% in order to keep electrostatic forces from building up on the surfaces of the particles and bed walls. A sensor (Sensirion AG, SHT31) measured the humidity at the exit of the humidifier. Despite the addition of humidity, the calculated density change of air is less than 1% and signatures of liquid bridging between the particles were not observed. The superficial gas velocity $U$ was close to the minimum fluidization velocity $U_{mf}$ of the bulk particles in the non-vibrated state. $U_{mf}$ was determined via pressure drop measurements between the top of the distributor plate and the freeboard of the bed. The values of $U$ ranged from $U/U_{mf}$ = [0.95, 1.04]. The combination of vibration and fluidizing air ensured that the particles were sufficiently mobilized while avoiding the formation of gas bubbles that could stir the granular bed and induce convective flow patterns [15].

A CCD camera (Photron GmbH, Fastcam SA Z.) with a framerate of 50 Hz was used to record the experiment. After recording, a digital image analysis was performed in order to track the position of the granular droplet and its daughter droplets. For this purpose, a MATLAB script tracked the leading edges of the granular daughter droplets and calculated their sinking speed. The tracking routine is shown in S1 in the SM [23]. Additionally, the script calculated the splitting angle of the granular daughter droplets, i.e. the angle enclosed by the two lines connecting the leading edges of the granular daughter droplets and the center of the initial position of the granular droplet (see inset of FIG. 6 (a)). The splitting angle was determined as a function of the distance travelled $y_h$, whereby $y_h$ is calculated as the mean vertical distance travelled by the left ($y_{hl}$) and right ($y_{hr}$) granular daughter droplet, see inset of FIG. 6 (a). For all of the recorded repetitions of an experiment, the function $\alpha(y_h) = k_1 e^{-k_2 y_h} + \alpha_{asym}$ was fitted to determine the asymptotic splitting angle $\alpha_{asym}$. A series of experiments were performed varying either the droplet size $D$, the airflow $U/U_{mf}$ or the vibration strength $\Gamma$. For each setting of the parameters $D$, $U/U_{mf}$ and $\Gamma$, three experiments were performed and averaged to calculate the sinking trajectory of the granular droplet and its fragmentation into daughter droplets. These experiments were complemented by particle image velocimetry (PIV) measurements allowing us to calculate velocity maps of the bulk particles. As tracer particles for PIV, 10% of the bulk particles were white glass particles with otherwise identical properties as the black bulk particles. The MATPIV 1.6.1. software [24] was used to calculate the velocity maps from consecutive image frames. The size of the PIV interrogation window was iteratively reduced from 64×64 to 32×32 pixels, using four iterations. The PIV data were then averaged over three frames i.e. a period of 0.3 s to reduce noise.

### B. Euler-Lagrange simulations of the granular droplet

#### 1. Modeling strategy: CFD-DEM

To numerically model the dynamics of granular droplets, a combined computational fluid dynamics and discrete element method (CFD-DEM) was used (open-source framework *CFDEM®Coupling*) [25]. CFD-DEM treats the particles as discrete entities (Lagrangian approach) that are immersed in a finite volume description of the fluid phase providing detailed information on the trajectories of each particle as well as inter-particle forces [26-30].

The DEM approach has been introduced by Cundall & Strack [31] and applies Newton's second law of motion to calculate the trajectory of each particle through force and torque balances. For any given particle $i$ of mass $m_i$ the force balance reads

$$m_i \frac{\partial^2 \boldsymbol{x}_i}{\partial t^2} = \boldsymbol{F}_{c,i} + \boldsymbol{F}_{fp,i} + \boldsymbol{g} m_i, \qquad (1)$$

where $\boldsymbol{F}_{c,i}$ and $\boldsymbol{F}_{fp,i}$ represent, respectively, the sum of all contact forces and the sum of all fluid-particle interaction forces acting on the particle and $\boldsymbol{g}$ is the gravitational vector. In the present work, particles were modelled as perfect spheres with elastic and isotropic material properties. The coefficient of restitution $e$ and the coefficient of friction $\mu$ for both materials are given in TABLE I. Particle contacts were described using the soft-sphere approach, whereby the normal contact forces follow the Hertz-Mindlin law expressed by a non-linear spring-dashpot model [27,32]. The tangential contact forces were also modelled by a non-linear spring-dashpot model, but were limited by the Coulomb criterion [32]. The gas flowing through the particle bed gave rise to various fluid-particle interaction forces $\boldsymbol{F}_{\text{fp},i}$. This work considered three fluid interaction forces: viscous force, pressure force, and drag force. Viscous and pressure forces stem from global variations in the fluid stress tensor on a length scale that is significantly larger than the particle size, whereas the drag force is due to flow variations in the vicinity of the particles [33]. It has been shown that the drag force dominates particle fluidization and was calculated here by the Koch-Hill closure model [34,35]. Other fluid-particle interaction forces such as virtual mass force or Basset force are negligible as the density of the gas is small compared to the particle density.

TABLE I. Particle properties.

| Property | Symbol (unit) | Simulation | | Experiment | |
|---|---|---|---|---|---|
| | | bulk | droplet | bulk | droplet |
| Particle diameter | $d$ (mm) | 1.7 | 1.1 | 1.7 ± 0.06 | 1.1 ± 0.06 |
| Density | $\rho$ (kg/m³) | 2500 | 6000 | 2500 | 6000 |
| Young's modulus | $E$ (MPa) | 5 | 5 | 63 | 210 |
| Poisson ratio | $\nu$ | 0.2 | 0.3 | 0.2 | 0.3 |
| Coefficient of restitution | $e$ | 0.88 | 0.7 | 0.88 ± 0.09 | 0.7 ± 0.1 |
| Coefficient of friction | $\mu$ | 0.47 | 0.38 | 0.47 ± 0.01 | 0.38 ± 0.05 |
| Minimum fluidization velocity[‡] | $U_{\text{mf}}$ (m/s) | 1.11 | 1.22 | 0.92 ± 0.01 | 0.90 ± 0.04 |

Two-way coupling between the dispersed particles and the continuous gas phase was established through an "unresolved" CFD-DEM approach. Here, a finite volume method solves volume-averaged Navier-Stokes-equations for the gas phase on a numerical grid (CFD grid) with cell size slightly larger than the particle diameter, so that a no-slip boundary condition cannot be resolved on the particles, and thus a drag law must be used. This method accounts for the presence of the immersed particles by calculating the void fraction $\epsilon_{\text{f}}$ and a momentum exchange term $\boldsymbol{R}_{\text{pf}}$ in each CFD cell. $\epsilon_{\text{f}}$ is the volume fraction of a CFD cell that is occupied by the gas phase and $\boldsymbol{R}_{\text{pf}}$ is the sum of the drag forces $\boldsymbol{F}_{\text{d},i}$ acting on all particles that are located in a given CFD cell of volume $\Delta V$ [25,33,36]. Owing to its constant temperature and the low Mach number involved, an incompressible solver with constant gas density $\rho_{\text{f}}$ was used to model the gas flow, simplifying the volume-averaged Navier-Stokes-equations to [25]

---

[‡] In the text, $U_{\text{mf}}$ refers to the value of the minimum fluidization velocity of the bulk particles. The respective value of the droplet particles is given for sake of completeness.

$$\frac{\partial \epsilon_f}{\partial t} + \nabla \cdot (\epsilon_f \boldsymbol{u}_f) = 0, \qquad (2)$$

$$\frac{\partial (\rho_f \epsilon_f \boldsymbol{u}_f)}{\partial t} + \nabla \cdot (\epsilon_f \boldsymbol{u}_f \boldsymbol{u}_f) = -\epsilon_f \nabla p - \boldsymbol{R}_{pf} + \epsilon_f \nabla \cdot \boldsymbol{\tau}_f + \rho_f \epsilon_f \boldsymbol{g}, \qquad (3)$$

$$\boldsymbol{R}_{pf} = \frac{1}{\Delta V} \sum_{i=1}^{n} \boldsymbol{F}_{d,i}. \qquad (4)$$

The shear stress tensor $\boldsymbol{\tau}_f$ in Eq. (3) was calculated as for an incompressible Newtonian fluid [26]. To resolve the particle collisions and to ensure numerical stability, the DEM part used a Verlet-integration scheme with a time step of $10^{-5}$ s. As such a high temporal resolution was not required for the CFD part, the time step of the CFD solver was $10^{-4}$ s. Hence, every 10$^{th}$ DEM time step a phase coupling was performed. More details concerning the two-way coupling can be found in the literature [25].

## 2. Model setup

Setting up a simulation involved two steps: (i) the initialization of the DEM packing and (ii) the initialization of the CFD domain. For the DEM part, a box of width 400 mm, transverse thickness 10 mm and height 600 mm was created representing the sidewalls of the fluidized bed. Bulk particles were inserted from the top of this box to build up a random packing of 480 mm height. A square cuboid volume (width 50 mm, height 50 mm, transverse thickness 10 mm) was removed from the bulk particle packing and filled with granular droplet particles to initialize, as in the experiments, a square shaped granular droplet of $D$ = 50 mm. Subsequently, additional bulk particles were added to create a flat bed surface at a height of 480 mm. TABLE I lists the properties of the particles used in the simulation. After initialization of the packing, the box was subjected to a vertical sinusoidal motion with $A$ = 1 mm and $f$ = 10 Hz. The collision properties of the sidewalls are identical to those of the bulk particles. To model the gas phase, a CFD grid of cubic voxels with 5 mm edge length was generated covering the same space as the DEM box. At the bottom of the CFD grid, gas was injected with a constant speed. The gas had a constant density and viscosity of 1.2 kg/m³ and 18 µPas, respectively, resembling properties of air. TABLE II summarizes all boundary conditions used for the CFD grid.

TABLE II. Boundary conditions used in the CFD part.

| Quantity | Symbol (unit) | inlet | outlet | wall |
|---|---|---|---|---|
| Reduced pressure | $p/\rho_f$ (m²/s²) | zero gradient | $10^5$ | zero gradient |
| Gas velocity | $\boldsymbol{u}_f$ (m/s) | $(0, U, 0)^T$ | zero gradient | full slip |
| Particle velocity | $\boldsymbol{u}_s$ (m/s) | zero gradient | zero gradient | zero gradient |
| Momentum coupling | $\boldsymbol{R}_{pf}$ (kg/(m³s)) | zero gradient | zero gradient | zero gradient |
| Void fraction | $\epsilon_f$ | 1.0 | 1.0 | zero gradient |

## III. RESULTS AND DISCUSSION

### 1. Fragmentation of a granular droplet

Both experiments and numerical simulations were applied to elucidate the underlying physics behind the phenomenon of granular droplet fragmentation. FIG. 1 (a) shows a comparison between experimental measurements and numerical simulations of the fragmentation dynamics of a granular droplet with $D = 50$ mm. Upon the start of vibration and fluidization, the square-shaped granular droplet flattens into a cap-shaped droplet (1.7 s). With time the cap-shaped droplet thins further, with the left and right "edges" penetrating the bulk phase, starting to form two daughter droplets that are, however, still connected by a thick band of particles (3.7 s). The daughter droplets descend along inclined trajectories with a similar inclination angle. During the descent, the connecting band between the two daughter droplets thins out and they eventually fully detach from each other (7.5 s). The entire process that leads to the formation of two separated daughter droplets is referred to as a binary split. After the detachment, each daughter droplet flattens and thins out, forming again two leading edges at each droplet (11.2 s) with each undergoing eventually another binary split and forming again two fragments for each daughter droplet. The series of binary splits continuous until the formed fragments are too small to split further or the bottom of the bed is reached. As can be seen from FIG. 1 (a), the numerical simulations agree well with the experimental data with regards to droplet shape and timescale of the splitting.

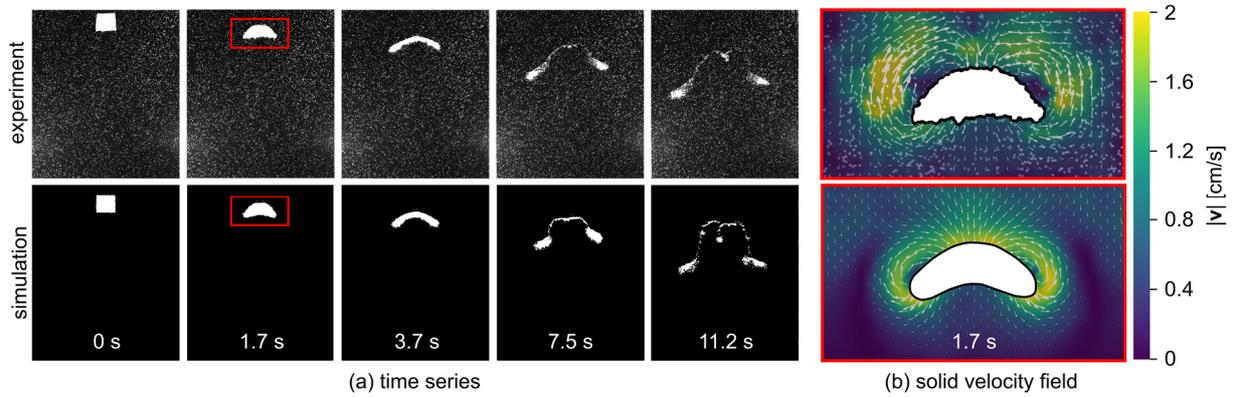

FIG. 1. Comparison between experimental measurements (top row) and simulation results (bottom row) on the dynamics of the splitting of a granular droplet ($D = 50$ mm) in vibro-fluidized bed using $f = 10$ Hz, $A = 1$ mm, and $U/U_{mf} = 0.95$ (exp.) or $U/U_{mf} = 0.97$ (sim.). (a) Time series showing the evolution of a granular droplet during fragmentation. Bulk particles are black, droplet particle are white (particle properties according to TABLE I). In the experiments, white tracer particles were added to the bulk phase for PIV measurements. (b) Velocity field of the bulk particles around the granular droplet as extracted from the red-framed region in (a) at 1.7 s. The granular droplet is white. The bulk particle phase is color-coded according to the velocity magnitude of the bulk particles. The velocity data was averaged over a period of 0.3 s to reduce noise in the PIV data.

In order to elucidate the underlying physics that drive the deformation and ultimately the splitting of the square-shaped granular droplet, the velocity field of the bulk particle phase is probed. FIG. 1 (b) shows the velocity field of the bulk particles in the vicinity of the granular droplet shown in FIG. 1 (a) at 1.7 s, as obtained by PIV measurements (top) and CFD-DEM simulations (bottom). In both cases, the velocity field of the bulk particles exhibits one vortex on each side of the granular droplet. At the center of the vortex, there is no granular movement. Overall, the bulk particles move from the leading edges to the roof of the granular droplet with the highest velocities (20 mm/s) being reached on a circular path with a radius of 20 mm from the vortex center. This motion pattern indicates that the displacement of bulk particles that drive the sinking motion of the droplet occurs primarily at the leading edges of the droplet. Due to conservation of mass, the displaced bulk particles move along circular paths towards the top of the droplet filling the space that has been occupied previously by the sinking droplet particles. In contrast to the rapid particle motion along the circular paths at the sides of the droplets, the bulk particle phase directly underneath the granular droplet appears to be largely immobile. This immobility is in contrast to what would be expected from an analogy of droplets falling in a liquid, where the displacement of the surrounding liquid is strongest underneath the sinking droplet. The immobility of this zone points to a frictional mechanism that locally reduces the fluidity of the bulk particle phase (immobilization) and obstructs the direct downward motion of the granular droplet. The existence of an immobilized zone also seems to be the main driver behind the splitting of the granular droplet. However, to support this hypothesis further insight into inter-particle contacts is required, which however is not accessible by PIV measurements and necessitates CFD-DEM data. Due to the good qualitative and quantitative agreement of the numerical and experimental results, see FIG. 1 (a) and FIG. 1 (b) respectively, CFD-DEM data was used to investigate the immobilized zone and the requirements for granular droplet fragmentation to occur.

### *2. Formation of an immobilized zone and the cause of fragmentation*

To understand better the formation of an immobilized zone and its effect on the granular droplet fragmentation, a CFD-DEM simulation was performed that tracked the particle motion with a temporal resolution of 2.5 ms. The high temporal resolution allows identifying different stages of particle motion within a single vibration cycle. In addition to the particle trajectories, the simulation provides information on the magnitude of the inter-particle forces for each particle contact. Here, the normal contact force $\mathbf{F}_{c,n}$ was used to localize force networks that constrain the motion of particles.

FIG. 2 shows snapshots of a granular droplet and the surrounding bulk particle phase for five consecutive stages of a vibration cycle. The first column visualizes the normal contact forces acting on each bulk particle by connecting contacting particle centers with a black line. The darker the line, the higher the magnitude of the normal contact force. The second column shows the particle velocity $v$ calculated in the reference frame of the vibrating bed, i.e. the $y$-component of $v$ is reduced by the instantaneous vertical velocity of the confining walls. The third column of FIG. 2 sketches the overall motion of the granular droplet (gray) and the bulk phase (green). Each row of FIG. 2 (a-e) corresponds to one particular stage (vibration phase φ) in a vibration cycle.

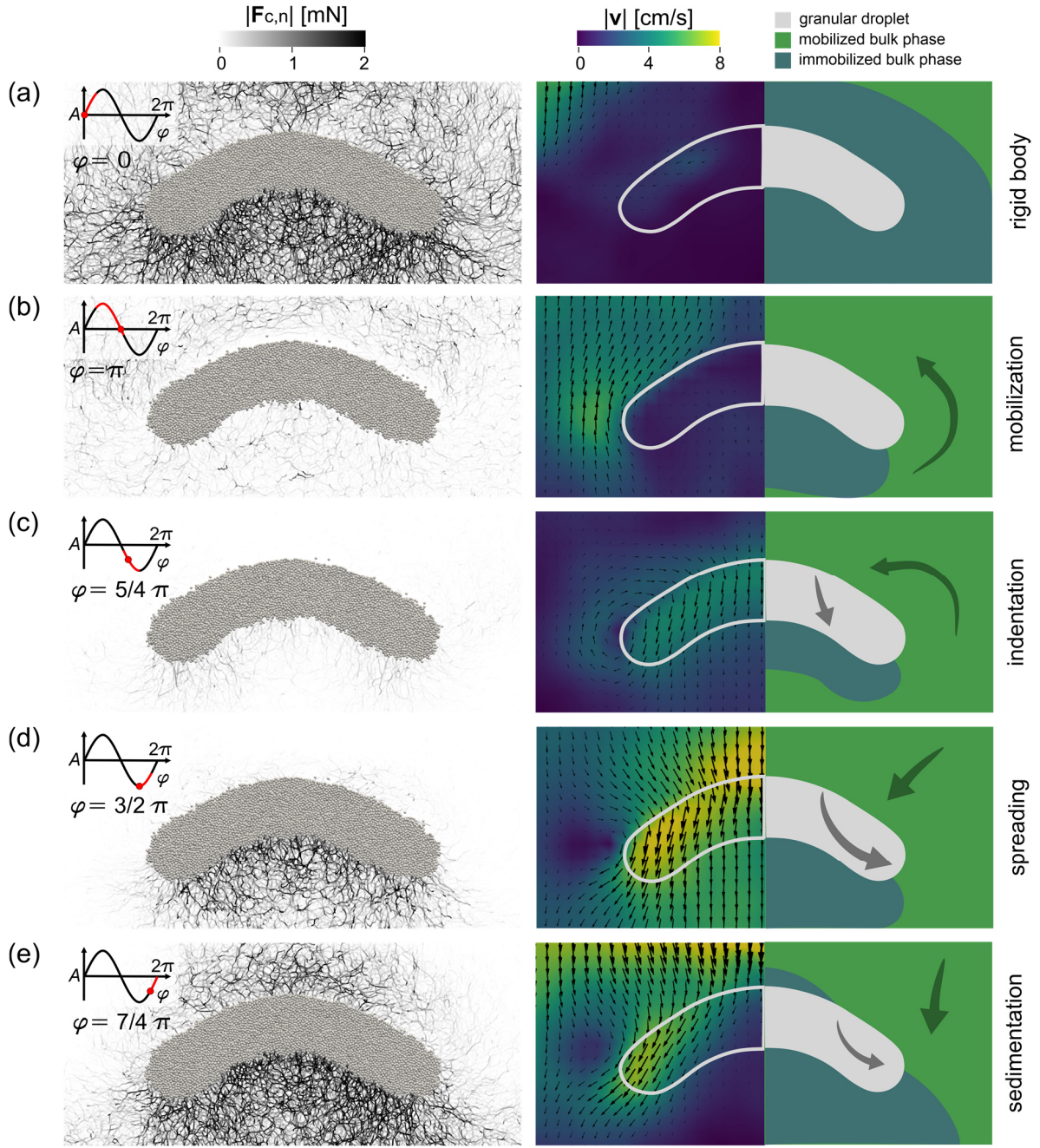

FIG. 2. The dynamics of the bulk and granular droplet particles at different stages of the vibration cycle (phase φ). Each row illustrates one of the five stages of motion identified with increasing φ from (a) to (e). The sketch in the left top corner of each row indicates the vertical displacement of the bed at the given phase (red dot) and the extent of the corresponding stage of motion (red part of the sine curve). In the first column, the droplet particles are gray and the normal contact force $\mathbf{F}_{c,n}$ acting between contacting bulk particles is shown as black lines. The opacity of the black lines increases with increasing contact force magnitude. In the second column, the granular droplet is shown as a gray contour and the particle velocity $v$ is visualized via color-coding and vector fields. Here, $v$ is calculated in the reference frame of the vibrating bed, i.e. the y-component of $v$ is reduced by the instantaneous vertical velocity of the confining walls. All results are obtained in a CFD-DEM simulation using $A = 1$ mm, $f = 10$ Hz, and $U/U_{\text{mf}} = 0.97$. The third column sketches the overall motion of the granular droplet (gray) and the bulk particle phase (green); light green indicates a mobilized and dark green an immobilized bulk phase.

Row (a) shows the vibration cycle at $\varphi = 0$, i.e. when the bed moves upwards, as indicated in the displacement sketch in the top left corner. Here, a network of strong inter-particle forces pervades almost the entire bulk phase, as can be seen in the contact force panel. This network inhibits any individual motion of the particles, hence the particles move in accordance with the imposed vibration of the bed with $v = 0$, i.e. the particle velocity relative to the wall velocity is zero as shown in the second column of row (a). This rigid body movement lasts from $\varphi = 0$ to $\pi/4$ up to the point when the mobilization of the bulk particles is initiated.

In the mobilization stage ($\varphi = \pi/4$ to $\pi$), the bed has passed the cusp of the sinusoidal vibration and moves downwards again. This downwards movement combined with the effect of the fluidizing gas causes a loosening of the bulk phase and the bulk particles above and at the sides of the granular droplet move upwards relative to the walls as illustrated in the particle velocity panel of row (b). Simultaneously, the loosening is accompanied by a reduction of the contact force network in the entire bed. In contrast to the mobilized bulk phase above and besides the granular droplet, the particle velocity $v$ within the granular droplet is almost zero, i.e. the granular droplet particles remain stationary relative to the walls. This is due to a smaller degree of fluidization of the droplet particles compared to the bulk particles (vide infra II.4), i.e. the droplet particles are not fully fluidized by the gas flow and the dominating gravity leads to the droplet following the movement of the walls (see also S2 in the SM [23]). Moreover, the stationary droplet particles inhibit the motion of the bulk particles directly below the granular droplet, which can be inferred from $v$ being comparably small in this region.

During the indentation stage ($\varphi = \pi$ to $3/2\pi$) illustrated in row (c) at $\varphi = 5/4\pi$, the bulk particles on top of the granular droplet exhibit nearly no inter-particle contact forces, indicative of a highly mobilized state. Since at $\varphi = 5/4\pi$, $v$ is small in this zone, the mobilization has been completed at this stage of the vibration cycle and the bulk particles do not move relative to the walls anymore. However, the granular droplet clearly exhibits a downward velocity $v$ and starts to indent the subjacent bulk phase as seen in the second column of row (c). This indentation is caused by two contributions: 1) At $\varphi = \pi$ the direction of the vibrational acceleration has inverted and points upwards. Thus, the imposed acceleration acts in the opposite direction of the downwards motion of the vibro-fluidized bed and inertially compresses the bed from the bottom. 2) Due to their reduced fluidization during the mobilization stage, the granular droplet particles gain a higher momentum than the surrounding bulk particles and thus start to penetrate the subjacent bulk phase as the bed decelerates, see second column of row (c).

In the subsequent spreading stage ($\varphi = 3/2\pi$ to $7/4\pi$), the granular droplet pushes down the bulk phase underneath the droplet and simultaneously compacts this region, forming a dense contact force network as shown in row (d). Due to this strong force network, the region below the granular droplet consolidates and forms an immobilized zone of growing size. The immobilized zone obstructs the further vertical penetration of the granular droplet and redirects the droplet particles to penetrate into less consolidated and still mobilized bulk phase regions at the sides of the granular droplet. This can be seen from the particle velocity fields in row (d) and (e), where the droplet particles exhibit considerable lateral motion. The lateral spreading of the granular droplet forms leading edges of the droplet that progress along inclined paths, which - at a later stage - will transform into daughter droplets. Lastly, in the sedimentation stage ($\varphi = 7/4\pi$ to $2\pi$) shown in row (e), the contact force network in the immobilized zone has strengthened further, leading to essentially stagnant particles in the region below the granular droplet. The propagation of the granular droplet particles occurs exclusively at its edges. Bulk particles in the mobilized regions at the sides and on top of the droplet move down and sediment as the vibro-fluidized bed moves up again. The emerging contact forces above the granular droplet provide further evidence for the solidification of the bulk phase due to sedimentation. Finally, the cycle is repeated by the bed reentering the rigid body motion stage shown in row (a). All five stages (a-e) repeat in each vibration cycle leading to the spreading and ultimately splitting of a granular droplet.

### 3. Second binary split

The previous section has demonstrated that underneath a granular droplet there is an immobilized zone characterized by particles with a strong contact force network. This immobilized zone obstructs the direct sinking of the droplet particles and forces the granular droplet to spread and ultimately to split, leading to the formation of two daughter droplets. However, we also observed that the daughter droplets themselves can be subject to another binary split. The data shown in FIG. 3 provides further insight into the mechanism at play during the second split. The top row in FIG. 3 visualizes the velocity of the bulk particles $v$, near the right daughter droplet in the mobilization stage ($\varphi = \pi$) of a vibration cycle. The bottom row in FIG. 3 displays the density and strength of the normal contact forces acting between the bulk particles in the spreading stage ($\varphi = 3/2\pi$) of the vibration cycle. Each row illustrates the temporal evolution of the daughter droplet in the same phase of the vibration cycle. In frames 5.05 s and 7.55 s, the velocity of the bulk particles underneath the spreading granular droplet is lower than the velocity of the mobilized particles above the granular droplet. This implies that although thin, the connecting band of particles between the daughter droplets maintains a stationary zone underneath the center of the connecting band and prevents this zone from mobilizing. The contact force networks in the corresponding frames 5.08 s and 7.58 s demonstrate the existence of strong contact forces in the region underneath the connecting band. Particles in the immobilized zones are strongly consolidated such that the droplet particles contained in the right daughter droplet can only penetrate into the mobilized bulk phase to their right. Analogously, the left daughter droplet particles can only penetrate to their left. Comparison of the first three frames in FIG. 3 (5.05 s, 7.55 s and 10.55 s) reveals that the laterally advancing daughter droplets deplete the connecting band between the two daughter droplets until it ultimately thins out and the connection "ruptures". Simultaneously with the thinning of the connecting band, the velocity of the bulk particles underneath the connection band increases appreciably (see 7.55 s and 10.55 s). The corresponding frames in the bottom row (7.58 s and 10.58 s) show that the strong contact force network underneath the connecting band disappears once the connecting band between the daughter droplets is only a few particles thick or has partially ruptured while networks of dense contact forces prevail directly underneath the granular daughter droplet that has detached. Combining the velocity and contact force information, we observed that once the connecting region between the two daughter droplets is thinner than 10 droplet particles (10.55 s), it can no longer suppress the mobilization of the bulk particles underneath, and the immobilized zone is limited to the region directly below the forming daughter droplet. After the formation of the new, individual daughter droplets (10.85 s), the same particle flow pattern as the one sketched in FIG. 2 (b) forms around each detached daughter droplet, i.e. the bulk particles at the outer side of each of the daughter droplet mobilize and a strong contact force network is only maintained underneath each daughter droplet but not on its sides (see 10.58 s). Thus, the daughter droplets start to spread towards both sides, reshaping themselves and eventually leading to another binary split of the daughter droplet as seen in the frames 17.55 s and 20.55 s.

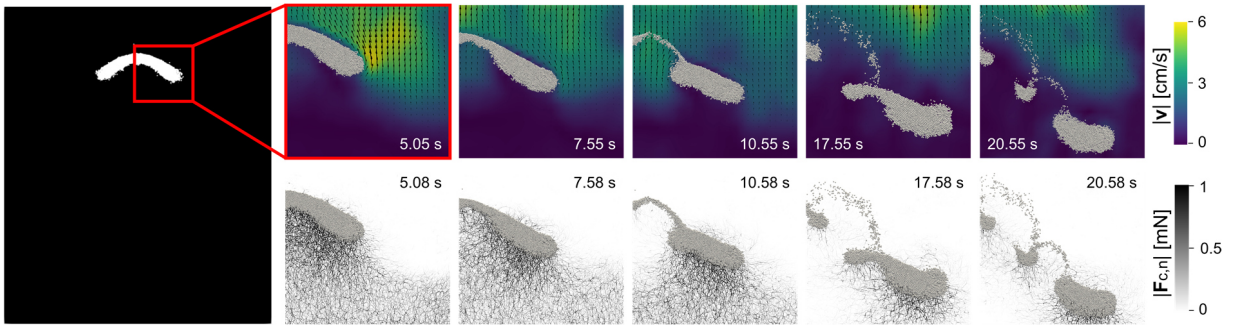

FIG. 3. Detachment process of the right daughter droplet and the initiation of a secondary split. The droplet particles are shown in gray; frames in the top row show the particle velocity field at the mobilization stage ($\varphi = \pi$), whereas the frames at the bottom row show the normal contact forces between the bulk particles and are taken in the corresponding spreading stage ($\varphi = 3/2\pi$). Each frame shows the same phase in the vibration cycle. The simulation was performed using $A = 1$ mm, $f = 10$ Hz and $U/U_{mf} = 0.92$. $U/U_{mf}$ is smaller than in FIG. 2 to make the second binary split to occur faster.

### 4. Particle requirements for granular droplet fragmentation to occur

In this section, we explore to what extent particle properties such as the coefficient of friction, the particle diameter, and the particle density influence the fragmentation of a granular droplet and the formation of daughter droplets. As these properties are difficult to control experimentally, a numerical parameter study was performed.

A systematic study of the effect of the coefficient of friction $\mu$ is presented in S3 in the SM [23]. FIG. 4 summarizes the key observations by comparing a reference case using a coefficient of friction $\mu = 0.3$ (top row) with a case using $\mu = 0$ (bottom row). In both cases, $\mu$ is identical for the bulk and droplet particles. In FIG. 4 (a), the time series for $\mu = 0.3$ shows how the initialized square droplet deforms to a cap-shaped droplet, leading to a continuously thinning connection between the two daughter droplets. As already visualized in FIG. 1 (b), also FIG. 4 (b) shows the presence of an immobilized zone underneath the droplet and high solid velocities at the outer regions of the granular droplet allowing for a lateral spreading and ultimately a binary split of the droplet. In contrast to the familiar behavior of the droplet for $\mu = 0.3$, the granular droplet evolves very differently for $\mu = 0$. Here, the droplet deforms in an opposite fashion as for $\mu = 0.3$, yielding a U-shaped granular droplet. The droplet sinks unimpededly through the bulk phase with a high velocity until, after only 1.1 s, it reaches the bottom of the bed. While sinking, the droplet broadens to some extent whereby thin trails of particles shed from the outer corners into the bulk phase. Importantly for $\mu = 0$, the granular droplet does not undergo a binary split, but instead, the droplet behaves very similar to a large liquid droplet sinking in an immiscible liquid of smaller density; see [16,37-39] for comparison. The velocity field of the bulk phase for $\mu = 0$ in FIG. 4 (b) reveals the absence of an immobilized zone underneath the granular droplet and a high downward directed velocity at the top of the granular droplet, allowing the granular droplet to rapidly migrate downwards. The displaced bulk particles flow in a large vortex structure to the top of the droplet. The centers of these vortices are located at the trailing edges of the granular droplet. This liquid-like behavior of the granular droplet is explained by the fact that the dense and less fluidized particles in the granular droplet can easily push aside the lighter particles of the bulk phase due to the absence of inter-particle friction. As soon as inter-particle friction is introduced, the sinking speed of the granular droplet is reduced and binary droplet splits occur; see S3 in the SM [23]. Hence, we can conclude that inter-particle friction is critical for the formation of an immobilized zone underneath the droplet and for a binary split to occur.

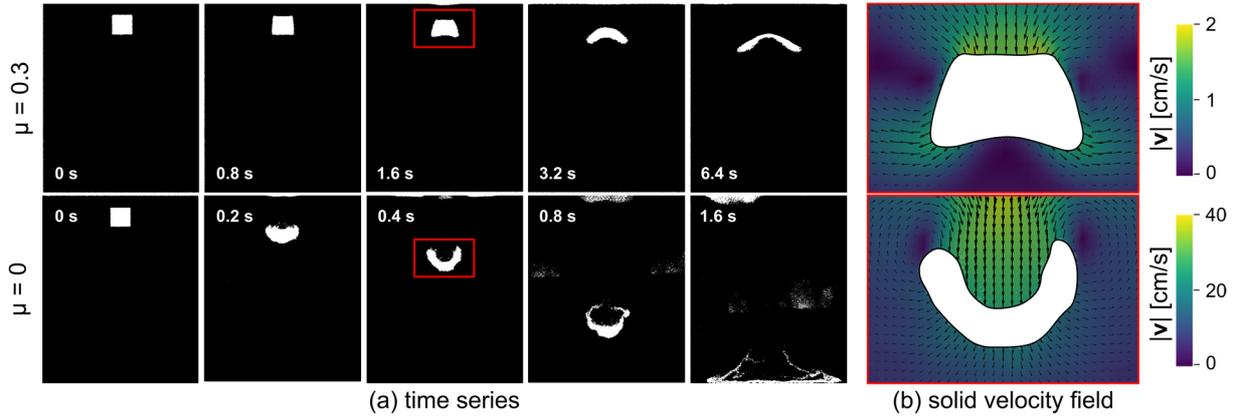

FIG. 4 Influence of the coefficient of friction $\mu$ on the sinking and fragmentation dynamics of a granular droplet of $D = 50$ mm. (a) Time series of granular droplet dynamics as obtained by CFD-DEM simulations using $A = 1$ mm, $f = 10$ Hz, $U/U_{mf} = 0.84$. Top row: simulation with a friction coefficient of $\mu = 0.3$; Bottom row: Friction coefficient $\mu = 0$. (b) Velocity of the bulk particles surrounding the granular droplet (white area) in the framed regions at $t = 1.6$ s (top row) and $t = 0.4$ s (bottom row). The background color quantifies the magnitude of the bulk particle velocity $|v|$. The velocity data are time averaged over 0.1 s. The simulation used a comparatively low value of $U/U_{mf}$ to reduce the emergence of large gas bubbles in the zero friction simulation.

The influence of particle density and particle size on the splitting behavior is shown in FIG. 5. This figure contrasts six simulations of varying relative particle size ($d_d/d_b$, three rows) and particle density ($\rho_d/\rho_b$, two columns). For each case, the position of the granular droplet and the gas flow field around the granular droplet are shown after a simulated time $t$. The gas flow field is quantified by the superficial velocity magnitude normalized by the inlet velocity $U$ of the vibro-fluidized. The gas velocity is averaged over one vibration period (0.1 s) to compensate for high-frequency gas fluctuations. FIG. 5 (a) shows the already known case of a granular droplet with particles of smaller size ($d_d/d_b$ = 0.645) and higher density ($\rho_d/\rho_b$ = 2.4) than the surrounding bulk particles (standard case as described in TABLE I). In the standard set-up the granular droplet sinks and splits into two daughter droplets. The corresponding gas velocity field displays the existence of a heterogeneous gas flow, where the gas velocity is higher around than within a granular droplet. In their original work, McLaren et al. [15] have argued that in case there is a size difference between the bulk particles and the particles in the granular droplet a heterogeneity in the gas flow is induced due to the lower permeability of the smaller granular droplet particles compared to the larger bulk particles. They further hypothesized this heterogeneity to be the major reason for the sinking and splitting of a granular droplet. To test this hypothesis, additional simulations varying $d_d$ and $\rho_d$ were assessed. For the case of equal particle size ($d_d/d_b$ = 1) and density ($\rho_d/\rho_b$ = 1), FIG. 5 (d), there is a homogenous bed permeability and thus a homogenous gas velocity across the bed. We further observed no coherent motion of the square-shaped initialized granular droplet over 15 s, but only diffusive mixing of the granular droplet particles with the surrounding bulk. The absence of any distinct motion of the granular droplet indicates that a granular droplet needs a driving size difference or density difference to display coherent motion. More importantly, this reference case confirms that the combination of vibration and gas flow applied does not introduce any large-scale convection patterns in the bed, which could otherwise influence the motion of the granular droplet. FIG. 5 (b) shows a simulation in which a granular droplet contains smaller particles, but of the same density as the surrounding bulk particles. Here, one observes a lower gas velocity in the granular droplet. However, despite this gas flow heterogeneity, the granular droplet does not show the expected sinking and splitting behavior, but instead rises. Thus, a heterogeneous gas flow alone does not result in a sinking and splitting granular droplet. In a further simulation, FIG. 5 (c), a case is modelled in which the particles in the granular droplet and the bulk have the same diameter, but the particles in the granular droplet have a higher density than the bulk particles. Since all particles have the same diameter, a fairly homogenous gas flow is established in the bed. Nevertheless, the granular droplet sinks, spreads and thins leading ultimately to a binary split. This finding indicates that density differences are the main driver for the splitting of a granular droplet. Only if the droplet particles are appreciably denser than the surrounding bulk particles, the region underneath the droplet can create an immobilized zone that is critical for droplet splitting to occur. To further reinforce this finding, FIG. 5 (e) and (f) show the results of two cases, where the droplet particles are larger than the bulk particles and the droplet particles are either denser than the bulk particles (e) or equally dense (f). In both cases, the gas flows preferentially through the granular droplet and the gas flow field is inverted to the ones found in (a) and (b). Despite this inversion, only in case (e) the granular droplet sinks and spreads laterally, whereas in (f) the droplet forms a plume that rises to the freeboard of the bed. A comparison of FIG. 5 (a, c, e) reveals a shorter time until the split of the granular droplet occurs with increasing $d_d/d_b$. A reason for the accelerated splitting dynamics could be that the heterogeneous gas flow around the droplet (in particular around its outer corners) helps to mobilize the surrounding bulk particles and facilitates the lateral spreading step discussed in FIG. 2 (c) and later in section III.5.

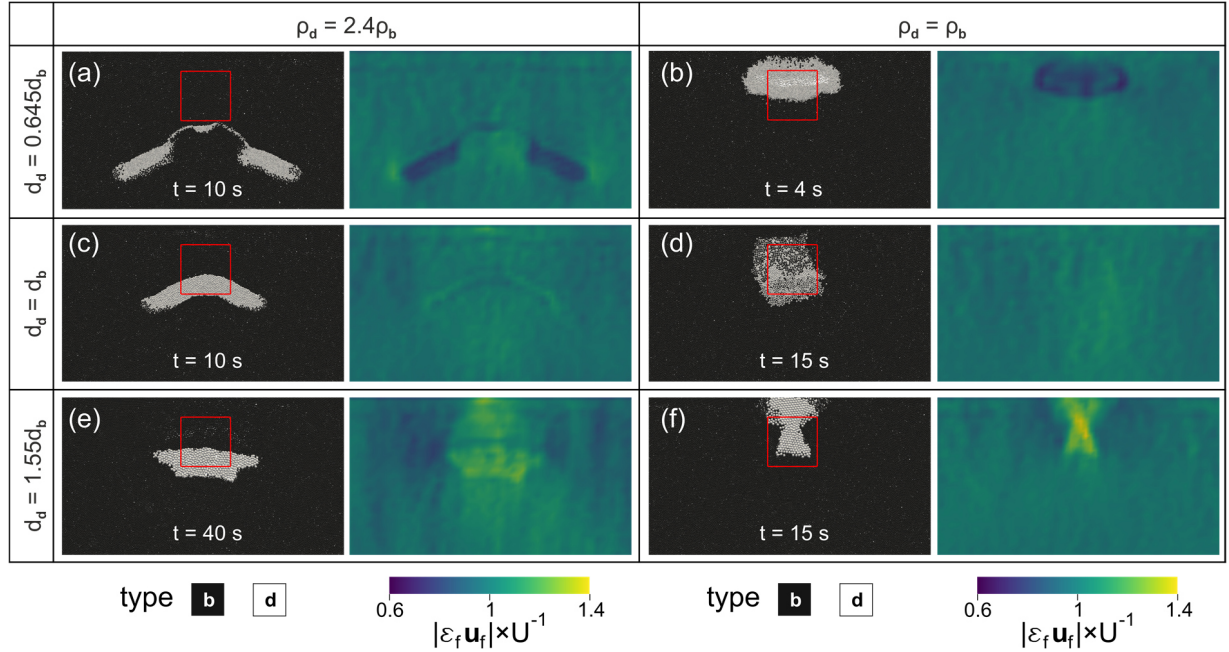

FIG. 5. Numerical study of the effect of the density ratio $\rho_d/\rho_b$ and size ratio $d_d/d_b$ between the droplet and the bulk particles on the granular droplet dynamics. Within a row, $d_d/d_b$ is constant (top to bottom: 0.645, 1, 1.55) and within a column $\rho_d/\rho_b$ is constant (left to right: 2.4, 1). In all simulations, $d_b = 1.7$ mm, $\rho_b = 2500$ kg/m³, $A = 1$ mm, $f = 10$ Hz and $U/U_{mf} = 0.92$. Each simulation result (a-f) is visualized in two panels: the left panel displays the instantaneous shape of the granular droplet (white) surrounded by the bulk particles (black) after a given time $t$. The red square marks the initial position of the granular droplet. The right panel shows the normalized superficial velocity of the gas flow $|\varepsilon_f \mathbf{u}_f|/U$ (color-coded) around the same granular droplet at $t$. The gas velocity data was averaged over one vibration period (0.1 s).

### 5. *Effect of vibration strength, fluidization level and droplet size on the splitting angle*

So far, we have established that the formation of an immobilized zone below the granular droplet is critical for droplet spreading and splitting and that this zone originates from frictional inter-particle contacts caused by a density difference between the bulk and droplet particles. Now, we further study the dynamics of a splitting granular droplet and how it is affected by changes in the granular droplet size $D$ and two external agitation parameters, i.e. the vibration strength $\Gamma$ and the fluidization level $U/U_{mf}$. As a quantification of the splitting dynamics, we use the splitting angle formed between the trajectories of two granular daughter droplets originating form a splitting event. It is expected that the extent of the immobilized zone will affect the trajectories of the daughter droplets, i.e. a larger extent of the immobilized zone should lead to a higher splitting angle $\alpha$. Here, we relied on an experimental approach, because the experiments were found to be less prone to gas bubbling than the numerical simulations.

The inset of FIG. 6 (a) shows the trajectories of a splitting granular droplet with $D$ = 30 mm using $U/U_{mf}$ = 1.01, $f$ = 30 Hz and $A$ = 0.05 mm. As can be seen from Fig. 6 (a), the initial value of α is close to 180° while it rapidly decreases with the spreading of the droplet. Once the two daughter droplets have detached from each other fully, α reaches an asymptotic value and the daughter droplets continue to descend at this asymptotic angle until they split again. The value of $α_{asym}$ obtained was subsequently used to qualitatively describe the extent of the immobilized zone that has built up underneath the daughter droplets. A high value of $α_{asym}$ corresponds to a large immobilized zone and vice versa. In FFIG. 6 (b-d) the values of $α_{asym}$ are plotted as a function of the vibration strength Γ, the fluidizing gas velocity $U/U_{mf}$, and the initial droplet size $D$.

When the vibration strength is increased, $α_{asym}$ decreases monotonically, as shown in FIG. 6 (b). Within the parameter range used in this work, the effect of the vibration strength on the asymptotic angle was the same (within a standard deviation) whether the amplitude or the frequency was changed, while the other was held constant, i.e. only the effective value of Γ is important for the dynamics of the daughter droplets. The assumed likely explanation is that an increase in vibration strength increases the energy input into the vibro-fluidized bed and increases the overall mobilization of the bulk particles, such that the granular droplet can penetrate the subjacent bulk phase more easily, leading to a decreasing $α_{asym}$. Further evidence of this facilitated vertical penetration is obtained when probing the sinking velocity of the daughter droplets and decomposing it into a vertical and horizontal velocity component, see FIG. 7. An increase in Γ accelerates the vertical sinking velocity, whereas the horizontal velocity of the daughter droplets, i.e. their lateral spreading velocity, is less affected. This, in turn, leads to an increase in the ratio of downward motion relative to the lateral motion of the daughter droplets and thus a smaller angle $α_{asym}$ as can be seen from FIG. 7 (a). Unlike the effect of Γ, an increase in $U/U_{mf}$ only leads to marginal changes in the asymptotic splitting angle within the precision of a standard deviation of the experiments, see FIG. 6 (b). The reason for this invariance is that an increase in $U/U_{mf}$ increases the mobility of both, the bulk phase beneath the granular droplet and especially the bulk phase at the leading edges of the splitting droplets due to gas flow heterogeneity. This can be seen in FIG. 7 (b), where an increase in $U/U_{mf}$ increases both the horizontal and vertical velocity of the daughter droplets. Next, the effect of the granular droplet size $D$ on the splitting angle is inspected. FIG. 6 (d) shows that an increasing size of the granular droplet leads to an increase in $α_{asym}$. This is due to an increasing size of the immobilized zone with increasing droplet size. In fact, we found that a minimum size for the granular droplet is required to observe the splitting of a droplet. A granular droplet with $D ≤ 10$ mm does not appear to form a sufficiently large, immobilized zone to give rise to droplet splitting; instead, a droplet of size $D ≤ 10$ mm simply sinks vertically without undergoing splitting, as can be seen in S4 in the SM [23]. This observation also explains why daughter droplets can only undergo a limited number of binary splits during their descent. Complementary numerical simulations have confirmed this minimum droplet size to yield droplet splitting.

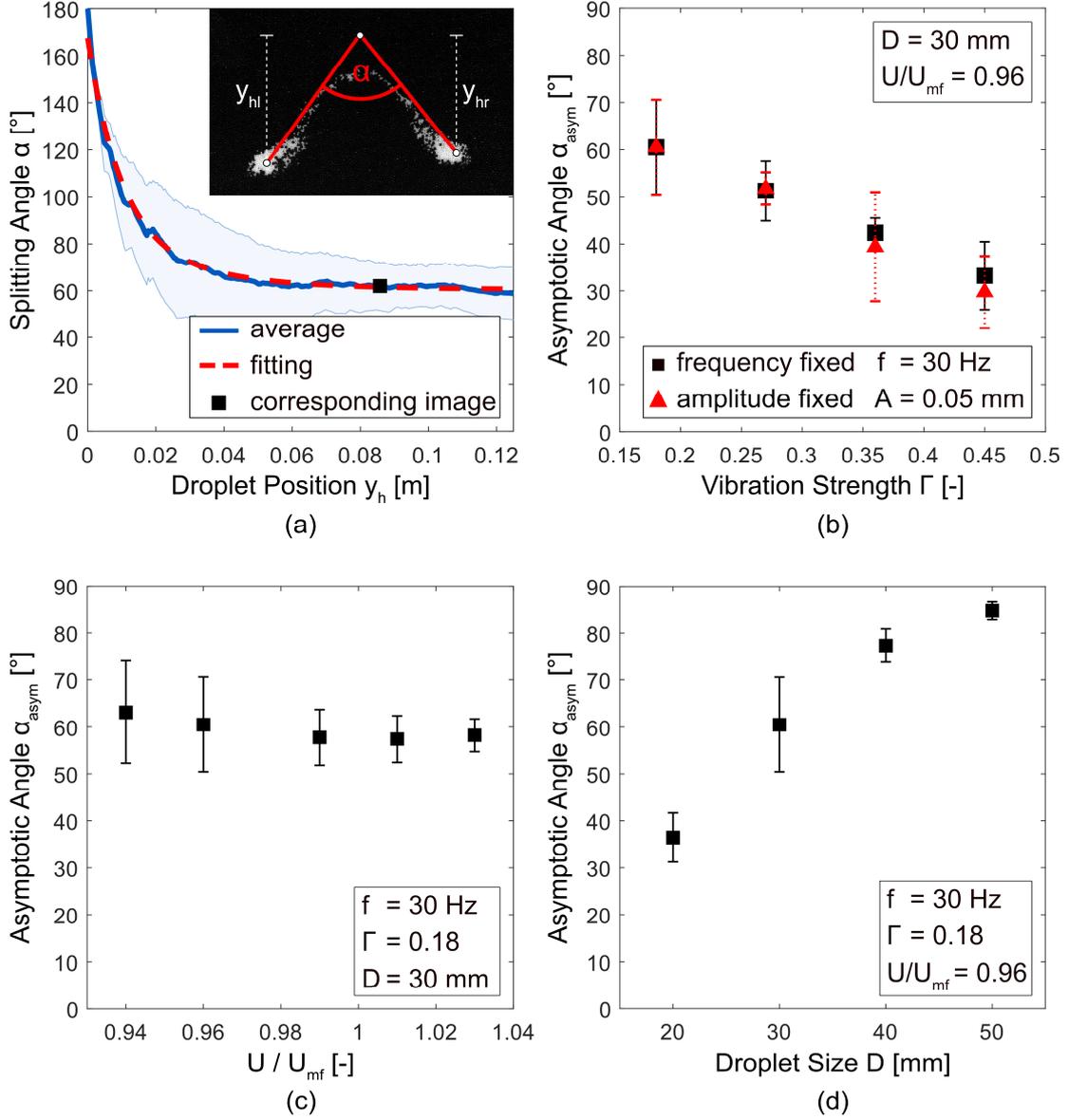

FIG. 6. Splitting angle as a function of vibration strength, fluidization level and droplet diameter: (a) Splitting angle α of a granular droplet as a function of vertical distance travelled $y_h$. The solid blue line represents the average splitting angle and the blue shaded area represents the standard deviation based on three repetitions. The dashed red line is fitted using $\alpha(y_h) = k_1 e^{-k_2 y_h} + \alpha_{asym}$. The inset shows the trajectories of the two daughter droplets evolving from a granular droplet of size $D$ = 30 mm for $U/U_{mf}$ = 0.98, $f$ = 30 Hz and $A$ = 0.05 mm. The red arc shows the construction of α at $y_h$ = 8.5 cm. (b) Effect of vibration strength on the asymptotic angle; the black squares plot $\alpha_{asym}$ for $D$ = 30 mm and $f$ = 30 Hz with varying $A$ and the red triangles plot $\alpha_{asym}$ for $D$ = 30 mm and $A$ = 0.05 mm for varying $f$. (c) Effect of $U/U_{mf}$ on $\alpha_{asym}$ for $D$ = 30 mm, $f$ = 30 Hz and $\Gamma$ = 0.18. (d) Effect of granular droplet size $D$ on $\alpha_{asym}$ for $f$ = 30 Hz, $\Gamma$ = 0.18 and $U/U_{mf}$ = 0.96. The error bars in (b), (c) and (d) represent the standard deviation based on three repetitions.

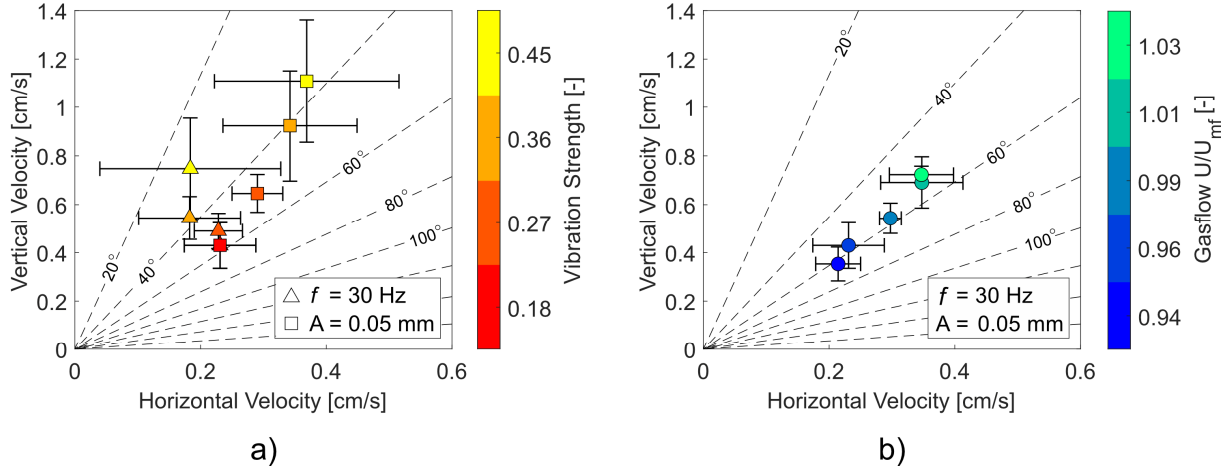

FIG. 7. Vertical and horizontal velocities of the sinking daughter droplets during their first split as a function of vibration strength (a) and gasflow (b). The results represent the averaged velocities of the leading edges as obtained by digital image analysis of the experiments (see S1 in [23]). The dashed lines in the background plot the asymptotic angle through $\alpha_{asym} = 2\tan^{-1}\left(\frac{v_{hor}}{v_{vert}}\right)$. The error bars show the standard deviation based on three experimental runs.

## IV. CONCLUSION

In this study, we have explored the physics behind the sinking, spreading and splitting of granular droplets in a vibro-fluidized bed. Although there are similarities with the Rayleigh-Taylor-induced fragmentation of liquid droplets falling in a miscible liquid of lower density, the mechanism causing the splitting of a granular droplet is different. We have demonstrated that the higher density of the particles forming the granular droplet compared to the density of the surrounding bulk particles leads to the formation of an immobilized region underneath the granular droplet that obstructs the direct downwards motion of the droplet. Instead, the droplet particles move preferably in the lateral direction, causing a spreading and thinning of the granular droplet and ultimately its splitting into two daughter droplets. The laterally moving daughter droplets stay initially connected, which maintains an immobilized zone underneath the elongated droplet. The two daughter droplets descend on inclined trajectories around this zone. Once the connecting band between the two forming daughter droplets becomes thinner than ten droplet particle diameters the weight of the connecting zone is insufficient to sustain further an immobilized zone. The disappearance of the immobilized zone initiates another binary split of each daughter droplet.

Numerical studies demonstrated further that a density difference between the bulk and the droplet particles is a key requirement for a splitting event to occur. Although size differences between the droplet particles and the bulk particles lead to a preferential flow of the gas through the region of the larger particles due to their higher permeability, this heterogeneous gas flow alone is not sufficient to cause a droplet to split. The formation of an immobilized zone further requires both inter-particle friction and a minimum droplet size greater than 10 mm; otherwise, the denser particles form a single droplet that sinks without fragmentation, similar to a droplet in an immiscible fluid. Additional experimental and numerical studies showed that the size of the immobilized zone increases with increasing initial droplet size and decreasing vibration strength, whereas $U/U_{mf}$ does not seem to extent the size of the immobilized zone appreciably.

The presented work has elucidated the physics driving the splitting of a granular droplet in a vibro-fluidized bed. However, there remains one open question: Why do the denser particles remain aggregated in the form of droplets instead of percolating through the bed as individual particles although a binary granular material does not possess any surface tension?

## V. ACKNOWLEDGMENT

This work was supported by the Swiss National Science Foundation Grant No. 200020 182692.

___________________________________________________________

# Supplemental Material to
# "The sinking dynamics and splitting of a granular droplet"

Jens P. Metzger,[1,*] Christopher P. McLaren,[1,*] Sebastian Pinzello,[1] Nicholas A. Conzelmann,[1] Christopher M. Boyce,[2] Christoph R. Müller,[1,†]

[1]*Department of Mechanical and Process Engineering, ETH Zurich, 8092 Zurich, Switzerland*
[2]*Department of Chemical Engineering, Columbia University, New York, New York 10027, USA*

[*] Shared first authors
[†] Corresponding author: muelchri@ethz.ch

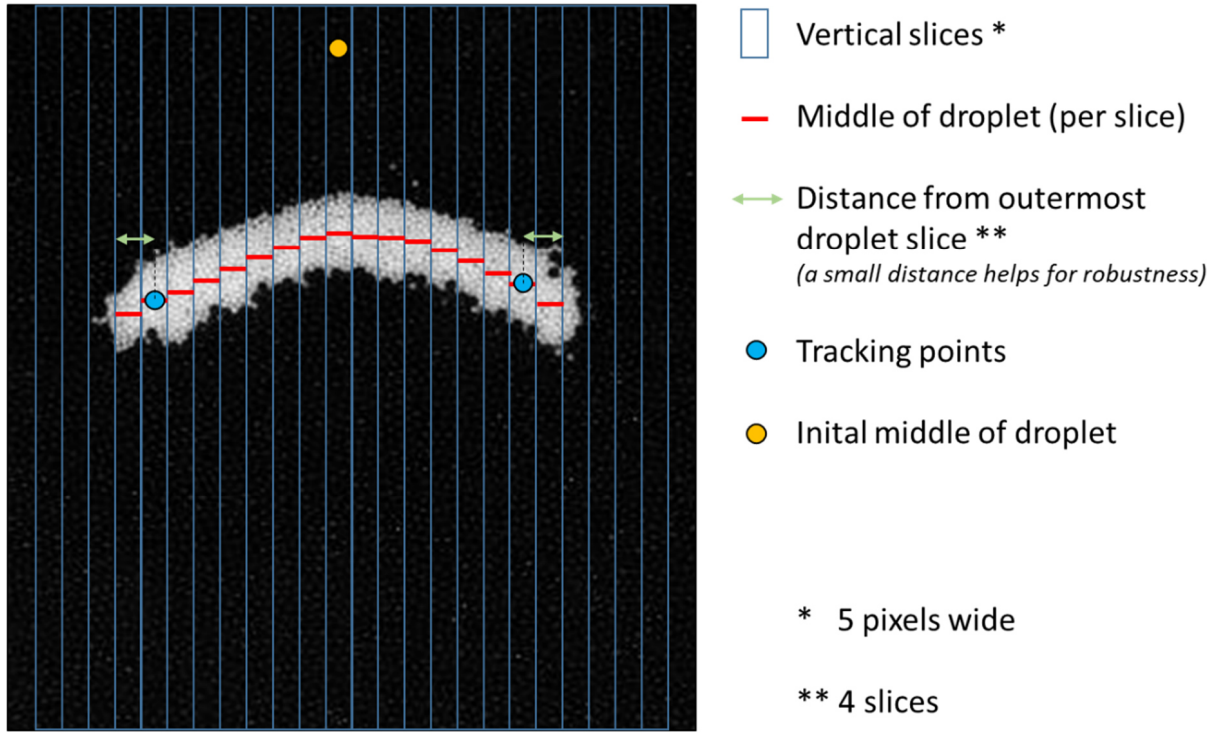

S1: Digital image analysis on a sinking granular droplet (white particles) about to form the two daughter droplets. The bed is divided into equally spaced vertical slices (each 5 pixels wide) to identify the leading edges of the daughter droplets formed from the initial granular droplet. For sake of tracking robustness, the tracking points (representing the leading edges) were set with a distance from the outermost slices that contain droplet particles.

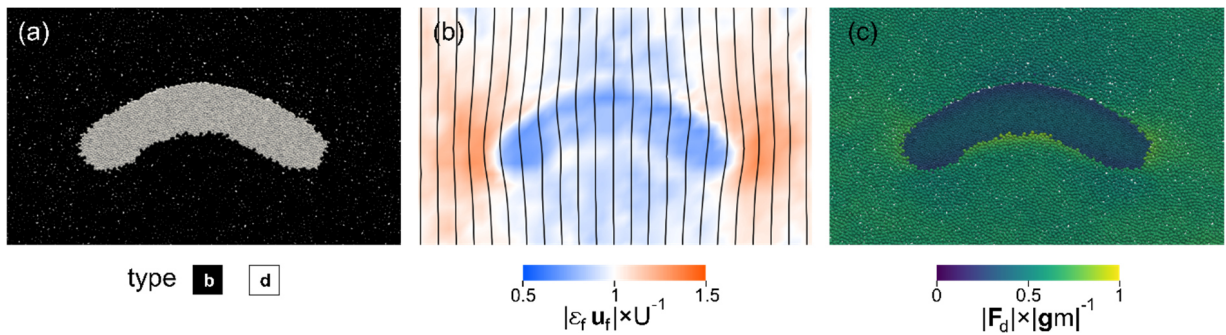

S2: (a) A granular droplet (white) spreads in the bulk phase (black) after t = 2.6 s of simulated time. (b) Gas flow field in the vicinity of a granular droplet. The magnitude of the superficial gas velocity $|\varepsilon \mathbf{u}_f|$ is averaged over two vibration periods (0.2 s) and normalized by the inlet velocity $U$ of the fluidized bed. Due to the lower permeability of the granular droplet compared to the bulk phase, the gas shifts around the droplet generating low flow velocities inside the granular droplet (blue field) and high flow velocities in the bulk phase left and right of the droplet (orange fields at the edges of the droplet). (c) Particle based drag force $\mathbf{F}_d$ normalized by the weight of the particle $|g m|$. The lack of gas flow inside the granular bubble reduces the drag force acting on the droplet particles and thus reduces the fluidization level $|\mathbf{F}_d| \times |g m|^{-1}$ compared to the bulk particles. The results are taken from simulations with the same settings as described in FIG. 2 (main article).

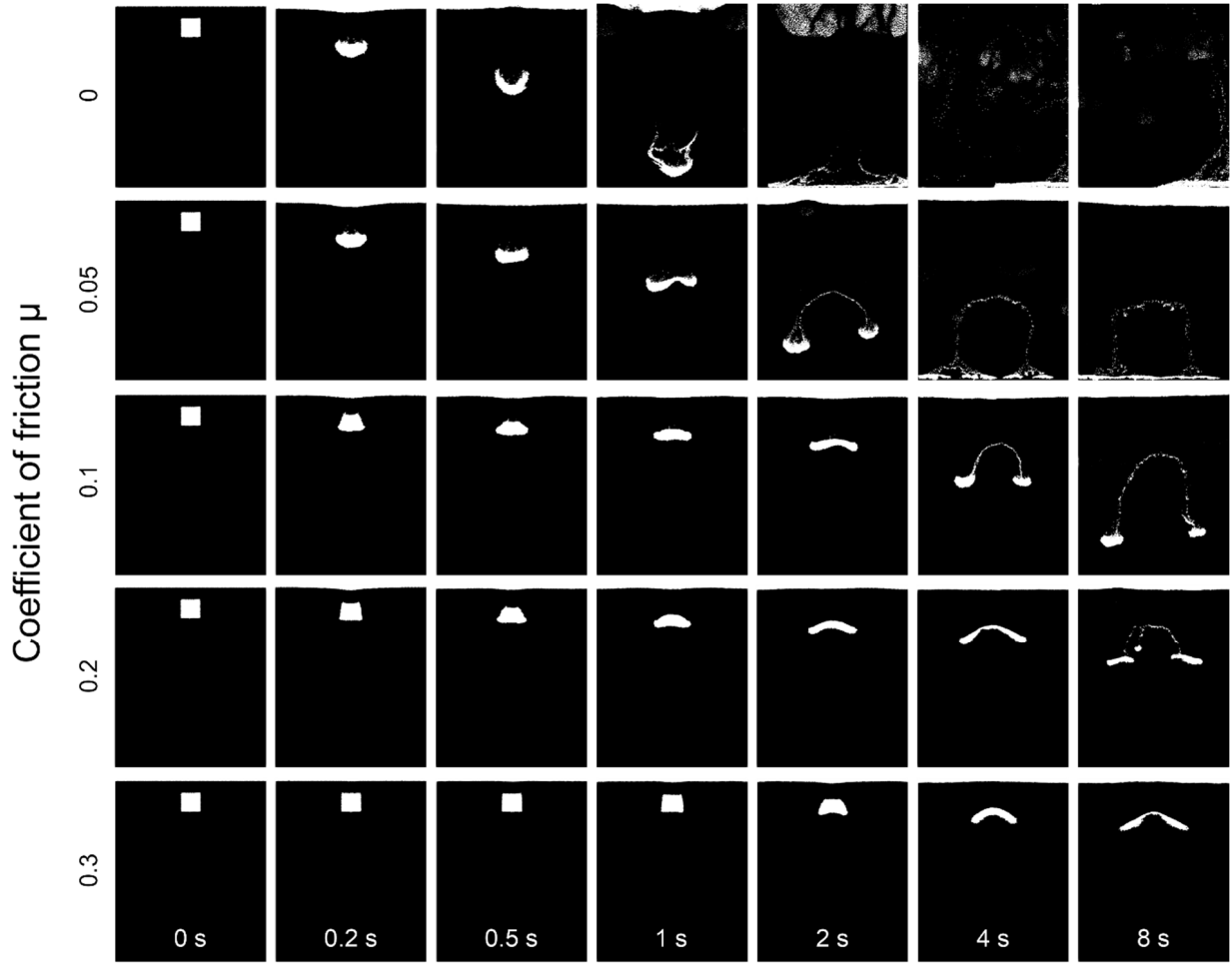

S3: Effect of the coefficient of friction $\mu$ on the temporal evolution of sinking granular droplets. Each row shows an individual time series (with increasing $\mu$ from top to bottom), where droplet particles (white) and bulk particles (black) share the same value of $\mu$. The results are obtained by CFD-DEM simulations with $A = 1$ mm, $f = 10$ Hz and $U/U_{mf} = 0.92$. For $\mu = 0$, gas bubbles emerged at the top half of the bed after 1 s of simulated time.

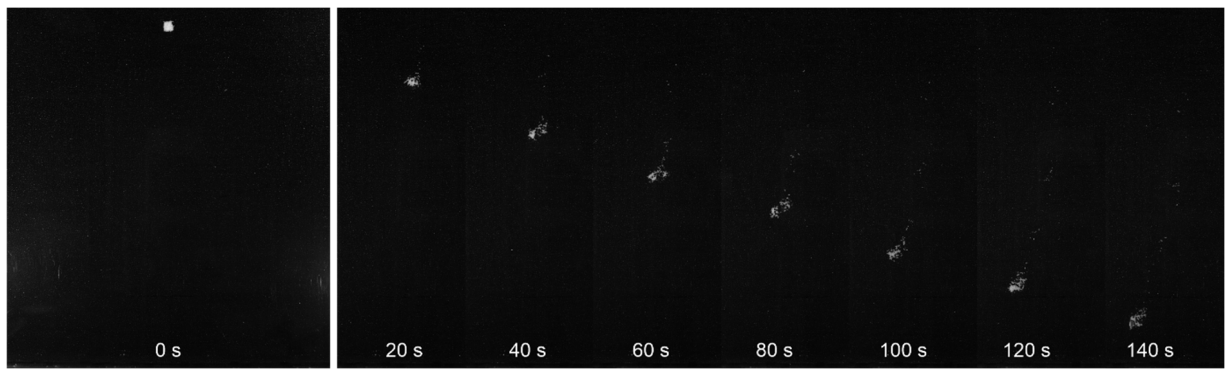

S4: Sinking of a granular droplet with $D = 10$ mm in a vibro-fluidized bed. The fluidizing gas flow is set to $U/U_{mf} = 0.96$ and the vibration is set to $f = 30$ Hz and $A = 0.05$ mm. Results are obtained by experiments. Particle densities and particle sizes of the granular droplet (white) and the bulk phase (black) are according to TABLE I (main article).

3